\documentclass[a4paper,10pt]{article}

\usepackage[cp1252]{inputenc}
\usepackage[english]{babel}
\usepackage{amsmath,amsfonts,amssymb,amsthm,amstext}
\usepackage{longtable,cite,url,graphicx}
\usepackage{color,relsize}

\newcommand\mysectiona[2]{%
\section
[#1]
{#2}
}

\newcommand\mysectionb[2]{%
\subsection
[#1]
{#2}
}

\newcommand\mysectionc[2]{%
\subsubsection
[#1]
{#2}
}

\newcommand{\notion}[1]{{\bf #1}}

\newcommand{\bigo}{\mathsf{O}}

\newcommand{\stdsetnameinform}[1]{%
\mathbb{#1}}

\newtheorem{definition}{Definition}

\newtheorem{theorem}{Theorem}

\newtheorem*{notation*}{Notation}
\newtheorem*{definition*}{Definition}
\newtheorem*{proposition*}{Proposition}
\newtheorem*{lemma*}{Lemma}
\newtheorem*{theorem*}{Theorem}
\newtheorem*{corollary*}{Corollary}

\begin{document}


%
\title{
\notion{FP//LINSPACE} computability of\\
Riemann zeta function $\zeta(s)$\\
in Ko--Friedman model 
}
\author{
Sergey\,V.\,Yakhontov
\date{}}

{\def\thefootnote{}
\footnotetext{
\centering
Sergey\,V.\,Yakhontov: Ph.D. in Theoretical Computer Science, Dept. of Computer Science, Faculty of\\
Mathematics and Mechanics, Saint Petersburg State University, Saint Petersburg, Russian Federation, 198504;\\
e-mail: SergeyV.Yakhontov@gmail.com, S.Yakhontov@spbu.ru; phone: +7-911-966-84-30;\\
personal Web page: \url{https://sites.google.com/site/sergeyvyakhontov/; 16-Nov-2014}}
%
\maketitle


\abstract{
%
In the present paper, we construct an algorithm for the evaluation of real Riemann zeta function $\zeta(s)$
for all real $s$, $s>1$, in polynomial time and linear space on Turing machines in Ko--Friedman model.
The algorithms is based on a series expansion of real Riemann zeta function $\zeta(s)$ (the series
globally convergents) and uses algorithms for the evaluation of real function $(1+x)^h$ and
hypergeometric series in polynomial time and linear space.

The algorithm from the present paper modified in an obvious way to work with the complex
numbers can be used to evaluate complex Riemann zeta function $\zeta(s)$ for $s=\sigma+\mathbf{i}t$,
$\sigma\ne 1$ (so, also for the case of $\sigma<1$), in polynomial time and linear space in $n$
wherein $2^{-n}$ is a precision of the computation; the modified algorithm will be also
polynomial time and linear space in $\lceil \log_2(t)\rceil$ and exponential time
and exponential space in $\lceil \log_2(\sigma)\rceil$.
%
}

\vspace{0.2cm}
%

\tableofcontents

%
\mysectiona
{Introduction}
{Introduction}
In the present paper, we consider computable real numbers and functions that are
represented by Cauchy functions computable by Turing machines \cite{K91} (Ko--Friedman model
of computable numbers and functions).

Main results regarding computable real numbers and functions can be found in \cite{A01,K91,K84,W00};
main results regarding computational complexity of computations on Turing machines can be found in
\cite{DK00}.

It is known that real Riemann zeta function $\zeta(s)$ \cite{E01} is a polynomial-time computable
real function when $s$ is a natural numbers \cite{K95,K98}; the algorithm from \cite{K95,K98} requires
at least $\bigo(n\log_2(n))$ memory cells to evaluate approximations of function $\zeta(s)$
to precision $2^{-n}$. Also, there is an algorithm for the evaluation of hypergeometric series
to high precision in quasi-linear time and linear space \cite{CGKZ05} which is applicable also for the
evaluation of $\zeta(3)$. The time and space complexity of the algorithms from \cite{K95,K98,CGKZ05}
are considered in the context of bit complexity; on Turing machines, the algorithm from \cite{K95,K98}
is polynomial in time and quasi-linear in space, and the algorithm from \cite{CGKZ05} is polynomial
in time and linear in space.

In the present paper, it is shown that real Riemann zeta function $\zeta(s)$ is polynomial-time and
linear-space (by the same algorithm) computable on Turing machines for all real $s$, $s>1$,
in Ko--Friedman model \cite{K91}. To prove that, we construct an algorithm for the evaluation of
real Riemann zeta function $\zeta(s)$ in polynomial time and linear space on Turing machines.
The algorithms is based on a series expansion from \cite{H30,S94} (the series globally convergents)
of function $\zeta(s)$ for complex $s$, $s\ne 1+2\pi \mathbf{i} \frac{n}{\log_2(2)}$, and uses
algorithms from \cite{Y08,Y10,YKK12} for the evaluation of real function $(1+x)^h$ and hypergeometric
series in polynomial time and linear space.

To derive the results, forward and backward error analysis of numeric algorithms \cite{H96} is
used in the preset paper. The method of such analysis for series is similar to one in
\cite{M93,Y08,Y10,YKK12}; in some sense, the algorithm described in the present paper is an applying
of the evaluation of complex numeric series in polynomial time from \cite{M93,Y08,Y10,YKK12} to the
numeric series that are used for the evaluation of real Riemann zeta function $\zeta(s)$.
%
%
\mysectionb
{$CF$ computable real numbers and functions}
{$CF$ computable real numbers and functions}
Cauchy functions in the model defined in \cite{K91} are functions binary converging to real numbers.
A function $\phi:\mathbb{N}\rightarrow\mathbf{D}$ (here $\mathbf{D}$ is the set of dyadic rational
numbers) is said to binary converge to real number $x$ if $|\phi(n)-x|\leq 2^{-n}$ for all $n\in\mathbb{N}$;
$CF_x$ denotes the set of all functions binary converging to $x$.

\begin{definition}
{\normalfont\cite{K91}}
Real number $x$ is said to be a $CF$ computable real number if $CF_x$ contains a function $\phi$
that is computable on Turing machines.
\end{definition}
\begin{definition}
{\normalfont\cite{K91}}
Real function $f$ on interval $[a,b]$ is said to be a $CF$ computable function on interval $[a,b]$
if there exists a function-oracle Turing machine $M$ such that for all $x\in[a,b]$
and for all $\phi\in CF_{x}$ function $\psi$ computed by $M$ with oracle $\phi$
is in $CF_{f(x)}$.
\end{definition}
%
%
\mysectionb
{Computational complexity of real functions}
{Computational complexity of real functions}
\begin{definition}
{\normalfont\cite{K91}}
Function $f:[a,b]\rightarrow \stdsetnameinform{R}$ is said to be computable in time $t(n)$ real
function on interval $[a,b]$ if for all computable real numbers $x\in[a,b]$ function $\psi\in CF_{f(x)}$
{\normalfont(}$\psi$ is from the definition of $CF$ computable real function{\normalfont)} is
computable in time $t(n)$.
\end{definition}
\begin{definition}
{\normalfont\cite{K91}}
Function $f:[a,b]\rightarrow \stdsetnameinform{R}$ is said to be computable in space $s(n)$ real
function on interval $[a,b]$ if for all computable real numbers $x\in[a,b]$ function $\psi\in CF_{f(x)}$
{\normalfont(}$\psi$ is from the definition of $CF$ computable real function\normalfont{)} is
computable in space $s(n)$.
\end{definition}
The input of functions $\phi$ and $\psi$ is $0^n$ ($0$ repeated $n$ times) when
a number or a function is evaluated to precision $2^{-n}$.  

\notion{FP//LINSPACE} denotes the class of string functions computable in polynomial time and
linear space (by the same algorithm) on Turing machines. According to this notation,
polynomial-time and linear-space computable real functions are said to be \notion{FP//LINSPACE}
computable real functions. The set of \notion{FP//LINSPACE} computable real functions on interval
$[a,b]$ is denoted by \notion{FP//LINSPACE}$_{C[a,b]}$.
%
%
\mysectionb
{Evaluation of approximations of real functions}
{Evaluation of approximations of real functions}
\label{Sec:FuncsEval}
Let' use the following results from \cite{K84,Y08,Y10,YKK12} in the present paper.

To multiply $a$ by $b$ to precision $2^{-n}$ wherein $a$ and $b$ are real numbers such that
$a\le 2^p$ and $b\le 2^p$ for some natural number $p$, it is sufficient to evaluate $a$ and $b$ to
precision $2^{-m}$ for $m=L_{a\cdot b}(n,p)$ wherein $L_{a\cdot b}$ is a natural function that is
linear in its arguments.

To inverse $a$ to precision $2^{-n}$ wherein $a$ is a real numbers such that $a\ge 2^{-p}$ some natural
numbers $p$, it is sufficient to evaluate $a$ to precision $2^{-m}$ for $m=L_{1/a}(n,p)$ wherein
$L_{1/a}$ is a natural function that is linear in its arguments.

To evaluation function $(1+x)^h$ to precision $2^{-n}$ wherein real $x\in [2^{-p}-1,2^{p}-2]$ and
real $|h|<1$, it is sufficient to evaluate $x$ and $h$ to precision $2^{-m}$ for $m=L_{pow}^{(1)}(n,p)$
wherein $L_{pow}^{(1)}$ is a natural function that is linear in its arguments.

It can be shown that to evaluation function $(1+x)^h$ to precision $2^{-n}$ wherein real
$x\in [2^{-p_1}-1,2^{p_1}-2]$ and real $|h|<p_2$, it is sufficient to evaluate $h$ to precision
$2^{-m}$ for $m=L_{pow}^{(2)}(n,p_1,p_2)$ wherein $L_{pow}^{(2)}$ is a natural function that is linear in
its arguments.
%
%
\mysectiona
{\notion{FP//LINSPACE} evaluation of real function $\zeta(s)$}
{\notion{FP//LINSPACE} evaluation of real function $\zeta(s)$}
\label{Sec:RealZetaEval}
Let's consider a globally convergent series for complex Riemann zeta function, valid for all complex
numbers $s$ except $s=1+2\pi \mathbf{i} \frac{n}{\log_2(2)}$ for some integer $n$, from \cite{H30,S94}:
\begin{align}
\label{Def:ZetaSeries}
\zeta(s)=\frac{1}{1-2^{1-s}} \sum_{k=0}^\infty \frac {1}{2^{k+1}}
\sum_{q=0}^{k} (-1)^{q} {k \choose q} (q+1)^{-s}.
\end{align}
Let
\begin{align*}
&f(q,s)=(q+1)^{-s},\\
&g(k,q)={k \choose q},\\
&h(k,s)=\sum_{q=0}^{k} (-1)^{q} g(k,q) f(q,s),\\
&u(s)=\sum_{k=0}^\infty \frac {1}{2^{k+1}} h(k,s),\ \text{and}\\
&v(s)=\frac{1}{1-2^{1-s}};
\end{align*}
we write series \eqref{Def:ZetaSeries} as
\begin{align}
\label{Def:ZetaSeries1}
\zeta(s)=v(s) u(s).
\end{align}
Let $p$ be a natural number, $p\ge 1$; let real $\lambda=\log_2(1+2^{-p})$, and $s$ be a real
number such that $1+\lambda\le s\le 2^p$. Let's evaluate $\zeta(s)$ by equation \eqref{Def:ZetaSeries1} to
precision $2^{-n}$ wherein $n$ is a natural number for such $s$.

%
\mysectionb
{Evaluation of $v(s) u(s)$}
{Evaluation of $v(s) u(s)$}
Because
\begin{align}
\label{Eq:vsEstim1}
\begin{array}{l}
s\ge 1+\lambda,\\
2^{s}\ge 2^{1+\lambda},\\
2^{s-1}\ge 2^{\lambda},\\
2^{1-s}\le 2^{-\lambda},\ \text{and}\\
1-2^{1-s}\ge 1-2^{-\lambda},
\end{array}
\end{align}
we have
\begin{align*}
v(s)&=\frac{1}{1-2^{1-s}}\le \frac{1}{1-2^{-\lambda}}=\\
&=\frac{2^{\lambda}}{2^{\lambda}-1}=\frac{1+2^{-p}}{2^{-p}}=\\
&=1+2^{p}<2^{2p}.
\end{align*}
Further, we have
\begin{align*}
u(s)=(1-2^{1-s}) \zeta(s)\le C_1
\end{align*}
wherein $C_1$ is a constant (this fact follows from the estimations in paragraph \ref{Par:Urs}).

Therefore, to evaluate $\zeta(s)$ by equation \eqref{Def:ZetaSeries1} to precision $2^{-n}$,
it is sufficient to evaluate $v(s)$ and $u(s)$ to precision $2^{-n_1}$ wherein $n_1\ge n+C_2(p)$
($C_2(p)$ is a constant that depends on $p$).

To evaluate $v(s)$ to precision $2^{-n_1}$, let's use algorithm from \cite{Y08,Y10,YKK12} for
the evaluation of function $(1+x)^h$ to precision $2^{-n_1}$; for that, it is sufficient
to evaluate $s$ to precision $2^{-m_1}$ for $m_1=L_1(n_1,p)$ wherein $L_1$ is a natural
function that is linear in its arguments ($m_1$ is a natural number).

%
\mysectionb
{Evaluation of $u(s)$}
{Evaluation of $u(s)$}
Let's evaluate function $u(s)$ as follows:
\begin{enumerate}
\item[1)]
{evaluate $f(q,s)$ to precision $2^{-n_4}$;}
\item[2)]
{evaluate $g(k,q)$ to precision $2^{-n_4}$;}
\item[3)]
{evaluate $$h(k,s)^*=\sum_{q=0}^k (-1)^q g(k,q)^* f(q,s)^*$$ wherein $g(k,q)^*$ and $f(q,s)^*$ are
approximations of $g(k,q)$ and $f(q,s)$ accordingly to precision $2^{-n_4}$; let $2^{-n_3}$
is the precision of evaluation of $h(k,s)^*$;}
\item[4)]
{evaluate $$u_{\iota}(s)^*=\sum_{k=0}^{\iota} \frac {1}{2^{k+1}}h(k,s)^*;$$ let $2^{-n_2}$
is the precision of evaluation of $u_{\iota}(s)^*$.}
\end{enumerate}
To evaluate $f(q,s)$ to precision $2^{-n_4}$, let's use algorithm from \cite{YKK12} for
the evaluation of function $(1+x)^h$ to precision $2^{-n_4}$; for that, it is sufficient
to evaluate $s$ to precision $2^{-m_3}$ wherein $m_3=L_3(n_4,p,\lceil \log_2(q)\rceil)$ wherein
$L_3$ is a natural function that is linear in its arguments ($m_3$ is a natural number).

To evaluate $u_{\iota}(s)^*$ to precision $2^{-n_2}$ by the series summation, it is sufficient
ot evaluate $h(k,s)^*$ to precision $2^{-n_3}$ such that $n_3\ge n_2+1$.
%
%
\mysectionc
{Evaluation of $g(k,q)$}
{Evaluation of $g(k,q)$}
Let's write $g(k,q)^{-1}$ as follows:
\begin{align*}
g(k,q)^{-1}&=\omega(k,q)=\prod_{\tau=1}^{q}b_{\tau}=\\
&=\prod_{\tau=1}^{q}\frac{q-\tau+1}{k-\tau+1}.
\end{align*}
Let's evaluate $\omega(k,q)$ in a loop for $\tau\in [1..(q-1)]$; at each step of the loop,
let's evaluate $$\omega(k,\tau)^* b_{\tau+1}^{*}$$ ($\omega(k,1)^*=b_1$) wherein $\omega(k,\tau)^*$ is an
approximation of $\omega(k,\tau)$ to precision $\epsilon_{\tau}$, $b_{\tau+1}^{*}$ is an approximation
of $b_{\tau+1}$ to precision $\epsilon_{\tau}$; $\epsilon_{\tau}=2^{-m}<2^{-1}$ for a natural number
$m$. Let's round $$\omega(k,\tau)^{*} b_{\tau+1}^{*}$$ to precision $\epsilon_{\tau}$
by dropping the bits after binary point from $q$-th bit to the rightmost bit.

Using mathematical induction for $\tau\in [1..(q-1)]$, let's show that 
\begin{align*}
\epsilon_{\tau}< 2^{-3q + 2\tau}
\end{align*}
holds for each $\tau\in [1..(q-1)]$ if we set $\epsilon_1\le 2^{-3q}$.

Base case: $\tau=1$; in that case, we evaluate $\omega(k,1)^*$, which is equal to $b_1^*$,
to precision $\epsilon_1\le 2^{-3q+2}$.

Inductive step: let $|\omega(k,\tau)^*-\omega(k,\tau)|\le \epsilon_{\tau}$ for $\tau\in [1..(q-2)]$.
In that case,
\begin{align*}
|\omega(k,\tau+1)^*-&\omega(k,\tau+1)|\le\\
&\le|\omega(k,\tau)^{*} b_{\tau+1}^{*}-\omega(k,\tau)b_{\tau+1}|+\epsilon_{\tau}=\\
&=|\omega(k,\tau)^{*} b_{\tau+1}^{*}-\omega(k,\tau)^{*} b_{\tau+1}+\\
&\quad +\omega(k,\tau)^* b_{\tau+1}-\omega(k,\tau) b_{\tau+1}|+\epsilon_{\tau}\le\\
&\le|\omega(k,\tau)^{*} (b_{\tau+1}^{*}-b_{\tau+1})|+\\
&\quad +|b_{\tau+1}(\omega(k,\tau)^* -\omega(k,\tau))|+\epsilon_{\tau}<\\
&< \epsilon_{\tau}+\epsilon_{\tau}+\epsilon_{\tau}<\\
&< 4\epsilon_{\tau}
\end{align*}
(here we use estimation $\omega(k,\tau)\le 1$). So, the following holds:

\begin{align*}
\epsilon_{\tau+1}<2^{-3q+2\tau+2}=2^{-3q+2(\tau+1)};
\end{align*}
in particular, $\epsilon_{q}<2^{-q}$. It means that it is sufficient to set $k=n_{\omega}+1$ and
$\epsilon_1\le 2^{-3q}$ to evaluate $\omega(k,q)$ to precision $2^{-n_{\omega}}$.

And to evaluate $g(k,q)=\omega(k,q)^{-1}$ to precision $2^{-n_4}$, it is sufficient
to set $n_{\omega}=L_g(n_4)$, wherein $L_{g}$ is a natural function that is linear in
its argument, because $\omega(k,q)\ge 2^{-k}$.
%
%
\mysectionc
{Evaluation of $h(k,s)^*$}
{Evaluation of $h(k,s)^*$}
Let's evaluate $h(k,s)^*=\sum_{q=0}^k (-1)^q g(k,q)^* f(q,s)^*$ in a loop for $q\in[0..k]$.
Because
\begin{align*}
h(k,s)^*&=\sum_{q=0}^k (-1)^q g(k,q)^* f(q,s)^*=\\
&=\sum_{q=0}^k (-1)^q (g(k,q)+\epsilon_g) (f(q,s)+\epsilon_f),
\end{align*}
wherein $\epsilon_g\le 2^{-n_4}$ and $\epsilon_f\le 2^{-n_4}$, we have
\begin{align*}
h(k,s)^*&=h(k,s)+\epsilon_h=\\
&=h(k,s)+\sum_{q=0}^k (-1)^q (g(k,q)\epsilon_f + f(q,s)\epsilon_g + \epsilon_g\epsilon_f).
\end{align*}
Further, because
\begin{align*}
g(k,q)\le 2^{k}\quad \text{and}\quad f(q,s)\le 1,
\end{align*}
the following hold:
\begin{align*}
|\epsilon_h|\le (k+1) \left( 2^k 2^{-n_4} + 2^{-n_4} + 2^{-2 n_4}\right)< 2^{C_3 k+C_4} 2^{-n_4}.
\end{align*}
It means that if $n_4=C_3 k + C_4 - n_3$ and if $g(k,q)$ and $f(q,s)$ are evaluated to precision $2^{-n_4}$
then $h(k,s)^*$ is evaluated to precision $2^{-n_3}$. 

%
\mysectionc
{Evaluation of $u(s)$}
{Evaluation of $u(s)$}
\label{Par:Urs}
Let's find a sufficient precision of the evaluation of $u(s)$ using the following equation:
\begin{align*}
|\Delta(u; s)|&=|u_{\iota}(s)^*-u(s)|\le\\
&\le|u_{\iota}(s)^*-u_{\iota}(s)| + |u_{\iota}(s)-u(s)|
\end{align*}
wherein
\begin{align*}
&u_{\iota}(s)^*=\sum_{k=0}^{\iota} \frac {1}{2^{k+1}}h(k,s)^*\quad \text{and}\\
&u_{\iota}(s)=\sum_{k=0}^{\iota} \frac {1}{2^{k+1}}h(k,s).
\end{align*}
Because
\begin{align*}
|u_{\iota}(s)^*-u_{\iota}(s)|\le 2^{-n_2},
\end{align*}
the following estimation holds: 
\begin{align*}
|\Delta(u; s)|\le 2^{-n_2} + |u_{\iota}(s)-u(s)|.
\end{align*}
So, the rest is to estimate
\begin{align*}
R_{\iota}(s)=|u_{\iota}(s)-u(s)|=v(s)\sum_{k=\iota+1}^{\infty} \frac {1}{2^{k+1}} h(k,s).
\end{align*}
Let $k$ is an odd natural number and $k'=(k\bmod 2)+1$; let
\begin{align*}
D_{k,q}(s)=g(k,q)f(q,s)-g(k,k-q)f(k-q,s).
\end{align*}
Because $g(k,q)=g(k,k-q)$, we have
\begin{align*}
|D_{k,q}(s)|&=|g(k,q)f(q,s)-g(k,k-q)f(k-q,s)|=\\
&=\left|g(k,q)\left((q+1)^{-s}-(k-q+1)^{-s}\right)\right|=\\
&=|g(k,q) d_{k,q}(s)|.
\end{align*}
Further, because
\begin{align*}
&(q+1)^{-s}\le 1,\\
&(k-q+1)^{-s}\le 1,\ \text{and}\\
&g(k,q)<2^{k},
\end{align*}
the following holds:
\begin{align*}
&|d_{k,q}(s,s)|<1\quad \text{and}\\
&|h(k,s)|\le \sum_{q=0}^{k'} |D_{k,q}(s)|\le \frac{k+1}{2} 2^{k'}.
\end{align*}
(the same is in the case of $k$ is an even natural number).
So,
\begin{align*}
R_{\iota}(s)&=v(s)\sum_{k=\iota+1}^{\infty} \frac {1}{2^{k+1}} \cdot \frac{k+1}{2} 2^{k'}\le \\
&\le C_5 2^{2p} 2^{-2^{-1} \iota}.
\end{align*}
As a result, if we set $\iota=4 p+2 n_2+C_6$ then 
\begin{align*}
|\Delta(u; s)|&\le 2^{-n_2} + C_5 2^{2p} 2^{-2^{-1}(4p+2n_2+C_6)}\le\\
&\le 2^{-n_2}+2^{-n_2}=2^{-n_2+1}.
\end{align*}
It means that it is sufficient to set $n_2=n_2+1$ and $\iota=4 p+2 n_2+C_6$ to evaluate
$u(s)$ to precision $2^{-n_1}$.
%
%
%
%
%
%
%
%
%
%
%

%
\mysectiona
{Main result}
{Main result}
As a result, to evaluate $\zeta(s)$ to precision $2^{-n}$ for $s\in [1+\lambda,2^p]$,
it is sufficient
\begin{enumerate}
\item[1)]
{to evaluate $s$ to precision $2^{-m}$ for $m=L_s(n,p)$ wherein $L_s$ is
a natural function that is linear in its arguments, and}
\item[2)]
{to use $L_{prod}(n,p)$ multiplications on $L_s(n,p)$ digits numbers wherein $L_{prod}$ is
a natural function that is linear in its arguments.
}
\end{enumerate}
It means the following theorem holds.
\begin{theorem}
Real Riemann zeta function $\zeta(s)$ is in class \notion{FP//LINSPACE}$_{C[a,b]}$
for any interval $[a,b]$ such that $a=1+\lambda$ and $b=2^p$ wherein $\lambda=\log_2(1+2^{-p})$
and $p$ is a natural number, $p\ge 1$.
\end{theorem}
Taking into account section \ref{Sec:FuncsEval}, we conclude the following theorem holds.
\begin{theorem}
Real Riemann zeta function $\zeta(s)$ is computable in exponential time and exponential space
in $p$ for any interval $[a,b]$ such that $a=1+\lambda$ and $b=2^p$ wherein $\lambda=\log_2(1+2^{-p})$
and $p$ is a natural number, $p\ge 1$.
\end{theorem}
%
%
%
\mysectiona
{\notion{FP//LINSPACE} evaluation of complex function $\zeta(s)$ on lines $\sigma+\mathbf{i}t$}
{\notion{FP//LINSPACE} evaluation of complex function $\zeta(s)$ on lines $\sigma+\mathbf{i}t$}
It is known that complex Riemann zeta function $\zeta(s)$ is computable in time $t^{\epsilon}$ on lines
$\sigma+\mathbf{i}t$ \cite{OSch88} for any $\epsilon>1$ and fixed $\sigma$ (algorithm from
\cite{OSch88} uses a precomputation involving $\bigo(T^{\frac{1}{2}}+\epsilon)$ operations
wherein $t\le T+T^{\frac{1}{2}}$) and is computable in time $t^{C}$ on line $\frac{1}{2}+\mathbf{i}t$
\cite{H11}. It means both real and complex Riemann zeta function $\zeta(s)$ is exponential time
computable in $\lceil \log_2(t)\rceil$ using algorithms from \cite{OSch88,H11}.

The algorithm from the present paper modified in an obvious way to work with the complex
numbers can be used to evaluate complex Riemann zeta function $\zeta(s)$ for $s=\sigma+\mathbf{i}t$,
$\sigma\ne 1$ (so, also for the case of $\sigma<1$), in polynomial time and linear space in $n$
wherein $2^{-n}$ is a precision of the computation; the modified algorithm will be also
polynomial time and linear space in $\lceil \log_2(t)\rceil$ and exponential time
and exponential space in $\lceil \log_2(\sigma)\rceil$.

To show it we need to prove computability of complex function $f(s)=(1+x)^s$ on lines $s=\sigma+\mathbf{i}t$,
$\sigma\ne 1$, in polynomial time and linear space in $\lceil \log_2(t)\rceil$ because all the results
from section \ref{Sec:RealZetaEval} hold also for the evalutions with the complex numbers.

Let's use the following equation:
\begin{align*}
e^{x+\mathbf{i}y}=e^{x}\cdot e^{\mathbf{i}y}=e^{x}\left(\cos(y)+\mathbf{i}\cdot\sin(y)\right).
\end{align*}
There are algorithms in \cite{Y10,YKK12} for the evaluation of real functions $\sin(y)$ and $\cos(y)$
on arbitrary interval $[2^{-p},2^{p}]$, $p\ge 1$, in polynomial time and linear space in $p$; that
algorithms use additive reduction of interval (subtracting approximate value of $\pi$ to derive
approximate value of $y$ to be fit in an appropriate interval). Therefore, we can evaluate function
\begin{align*}
(1+x)^{s}=\exp(s\cdot \log(1+x))
\end{align*}
on arbitrary area $\sigma\in [1+\lambda,2^{p}]$ and $t\in [-2^{p},2^{p}]$, wherein $\lambda=\log_2(1+2^{-p})$
and $p\ge 1$, in polynomial time and linear space in $p$. So, complex function $(1+x)^s$ is computable
on lines $s=\sigma+\mathbf{i}t$, $\sigma>1$, in polynomial time and linear space in $\lceil \log_2(t)\rceil$;
therefore, it holds for complex Riemann zeta function $\zeta(s)$.

Now let's consider the evaluation of complex function $\zeta(s)$ on lines $\sigma+\mathbf{i}t$
for $\sigma\in [2^{-p},1-\lambda]$ and $t\in [-2^{p},2^{p}]$, wherein $\lambda=\log_2(1-2^{-p})$
and $p\ge 1$. 

Let's evaluate $\zeta(s)$ by equation \eqref{Def:ZetaSeries1} to precision $2^{-n}$ for such $s$
wherein $n$ is a natural number. Let's consider equation \eqref{Eq:vsEstim1} for complex $s$ and $\sigma<1$:
\begin{align*}
&|2^{s}|=|2^{\sigma+\mathbf{i}t}|=|2^{\sigma}|\ge |2^{-p}|,\\
&2^{s-1}\ge 2^{-p-1},\\
&2^{1-s}\le 2^{1+p},\ \text{and}\\
&1-2^{1-s}\ge 1-2^{1+p}
\end{align*}
(here $|\cdot|$ is the complex modulus);
therefore
\begin{align*}
|v(s)|&=\left|\frac{1}{1-2^{1-s}}\right|\le \frac{1}{1-2^{1+p}}.
\end{align*}
All the further estimations should be the same as for the case of real function $\zeta(s)$. 
%


\end{document}